\documentclass[useAMS,usenatbib,usegraphicx]{mn2e}
\usepackage{times}
\title[Non-Gaussian Signatures in Temperature Fluctuation Observed by WMAP]
  {Non-Gaussian Signatures in the Temperature Fluctuation Observed by the Wilkinson Microwave Anisotropy Probe}
\author[Chan-Gyung Park]
  {Chan-Gyung Park\thanks{E-mail: parkc@kias.re.kr} \\
Korea Institute for Advanced Study, 130-722, Korea}
\begin{document}


\pagerange{\pageref{firstpage}--\pageref{lastpage}} \pubyear{2003}

\maketitle

\label{firstpage}

\begin{abstract}
We present results from a test for the Gaussianity of the whole sky
sub-degree scale CMB temperature anisotropy measured by the Wilkinson 
Microwave Anisotropy Probe (WMAP). 
We calculate the genus from the foreground-subtracted and Kp0-masked WMAP maps 
and measure the genus shift parameters defined at negative and positive 
threshold levels ($\Delta\nu_{-}$ and $\Delta\nu_{+}$) and the asymmetry
parameter ($\Delta g$) to quantify the deviation from the Gaussian relation. 
At WMAP Q, V, and W bands, the genus and genus-related statistics imply that
the observed CMB sky is consistent with Gaussian random phase field. 
However, from the genus measurement on the Galactic northern and southern 
hemispheres, we have found two non-Gaussian signatures at the W band resolution
($0\fdg35$ scale), i.e., the large difference of genus amplitudes between 
the north and the south and the positive genus asymmetry in the south, 
which are statistically significant at $2.6\sigma$ and $2.4\sigma$ levels, 
respectively. The large genus amplitude difference also appears in the WMAP 
Q and V band maps, deviating the Gaussian prediction with a significance level 
of about $2\sigma$.
The probability that the genus curves show such a large genus amplitude 
difference exceeding the observed values at all Q, V, and W bands in a 
Gaussian sky is only 1.4\%. 
Such non-Gaussian features are reduced as the higher Galactic cut
is applied, but their dependence on the Galactic cut is weak.
We discuss possible sources that can induce such non-Gaussian features,
such as the Galactic foregrounds, the Integrated Sachs-Wolfe and the 
Sunyaev-Zel'dovich effects, and the reionisation-induced low $\ell$-mode 
non-Gaussianity that are aligned along the Galactic plane. 
We conclude that the CMB data with higher signal-to-noise ratio and 
the accurate foreground model are needed to understand the non-Gaussian 
signatures. 
\end{abstract}

\begin{keywords}
cosmology -- cosmic microwave background
\end{keywords}

\section{Introduction}

Recently the Wilkinson Microwave Anisotropy 
Probe\footnote{http://map.gsfc.nasa.gov} (WMAP) satellite mission 
has opened a new door to the precision cosmology. The WMAP has measured 
the cosmic microwave background (CMB) temperature anisotropy and polarisation 
with high resolution and sensitivity (Bennett et al. 2003a). 
The WMAP data implies that the observed CMB fluctuations are consistent 
with predictions of the concordance $\Lambda$CDM model with scale-invariant 
and adiabatic fluctuations which have been generated during the inflationary
epoch (Hinshaw et al. 2003; Kogut et al. 2003; Spergel et al. 2003; 
Page et al. 2003; Peiris et al. 2003).

An important feature of the simplest inflation models is that the primordial
density fluctuation field has a Gaussian random phase distribution 
(Guth \& Pi 1982; Hawking 1982; Starobinsky 1982; Bardeen, Steinhardt \& 
Turner 1983; see Riotto 2002 for a review). 
Therefore, an observational test of the Gaussianity of the initial density 
fluctuation field will provide an important constraint on inflation models. 
Fortunately, the CMB temperature anisotropy, which reflects the density 
fluctuation on the last scattering surface, is expected to be the best probe
of the primordial Gaussianity.

There have been many tests for the Gaussianity of the CMB anisotropy 
at large ($\sim 10\degr$; Kogut et al. 1996; Colley, Gott, \& Park 1996;
Ferreira, Magueijo, \& G\'orski 1998; Heavens 1998; Pando, Valls-Gabaud, 
\& Fang 1998; Novikov, Feldman, \& Shandarin 1999; Bromley \& Tegmark 1999; 
Magueijo 2000; Banday, Zaroubi, \& G\'orski 2000; 
Mukherjee, Hobson, \& Lasenby 2000; Barreiro et al. 2000; 
Aghanim, Forni, \& Bouchet 2001; Phillips \& Kogut 2001; Komatsu et al. 2002), 
intermediate ($\sim 1\degr$; Park et al. 2001; Shandarin et al. 2002), 
and small angular scales ($\sim 10\arcmin$; Wu et al. 2001; Santos et al. 2002,
2003; Polenta et al. 2002; De Troia et al. 2003). 
Most of the results implies that the primordial density fluctuation is 
consistent with Gaussian random phase. 
Recently, Komatsu et al. (2003) have presented limits to the amplitude of 
non-Gaussian primordial fluctuations in the WMAP 1-year CMB maps by measuring
the bispectrum and Minkowski functionals, and found that the WMAP data are 
consistent with Gaussian primordial fluctuations. 
Similar result has been obtained by Colley \& Gott (2003), who independently 
have measured the genus, one of the Minkowski functionals, from the WMAP maps.
Gazta\~naga \& Wagg (2003) and Gazta\~naga et al. (2003) also have concluded 
that the WMAP maps are consistent with Gaussian fluctuations from the 
measurement of the three-point correlation function and the higher-order 
moment of the two-point correlations, respectively. 
On the other hand, Chiang et al. (2003) argue that non-Gaussian 
signature has been detected from the foreground-cleaned WMAP map produced by 
Tegmark, de Oliveira-Costa, \& Hamilton (2003) using the phase mapping 
technique. However, it should be noted that their analysis has been done 
for the whole sky map whose Galactic plane region still has significant 
foreground contamination.

In this paper, we perform an independent test for the Gaussianity of 
the CMB anisotropy field using the WMAP 1-year maps by measuring the genus and 
the genus-related statistics. Previous works on the genus measurement from 
the WMAP simulation can be found in Park et al. (1998) and 
Park, Park, \& Ratra (2002, hereafter PPR). This paper is organized as follows. 
In $\S2$, we summarise how the recent release of the WMAP 1-year CMB sky maps
are reduced for data analysis. In $\S3$, we describe how the genus is measured
from the CMB maps, and give results from the WMAP 1-year maps. 
The genus for the north and the south hemispheres of the WMAP maps are compared
in $\S4$. Possible sources that can cause non-Gaussian 
features in the genus curve are discussed in $\S5$. 
We conclude in $\S6$.

\section{WMAP 1-year Maps}

The WMAP mission was designed to make the full sky CMB maps with high accuracy,
precision, and reliability. The sky map data products derived from 
the WMAP observations have 45 times the sensitivity and 33 times the angular 
resolution of the COBE/DMR mission (Bennett et al. 2003a). 
There are four W band ($\sim 94$ GHz), two V band ($\sim 61$ GHz), 
two Q band ($\sim 41$ GHz), one Ka band ($\sim 33$ GHz), and one K band 
($\sim 23$ GHz) differencing assemblies, with $0\fdg21$, $0\fdg33$, $0\fdg49$, 
$0\fdg62$, and $0\fdg82$ FWHM beam widths, respectively. 
The maps are prepared in the 
HEALPix\footnote{http://www.eso.org/science/healpix} 
format with ${\rm nside}=512$ (G\'orski, Hivon, \& Wandelt 1999). 
The total number of pixels of each map is $12\times {\rm nside}^2 = 3,145,728$.
Since the K and Ka band maps are dominated by Galactic foregrounds, 
we use only Q, V, and W band data in our analysis.

The Galactic dust, free-free, and synchrotron emissions are highly non-Gaussian
sources, and their contribution to the measured genus is not negligible 
even at high Galactic latitude (PPR). 
Therefore, we use the 100 $\mu{\rm m}$ dust (Schlegel, Finkbeiner, \& Davis 
1998; Finkbeiner, Davis, \& Schlegel 1999), H$\alpha$ (Finkbeiner 2003), and 
synchrotron (Haslam et al. 1981, 1982) emission maps multiplied with
coefficients given in Table 3 of Bennett et al. (2003b) to reduce 
the foreground emission in each differencing assembly map of the Q, V, 
and W bands. 
At each frequency band, we combine the foreground-subtracted WMAP differencing
assembly maps with noise weight given by $N_{\rm obs}/\sigma_0^2$, where 
$N_{\rm obs}$ is the effective number of observations at each pixel, 
and $\sigma_0$ is the global noise level of the map (Bennett et al 2003a, 
Table 1). 
Finally, we obtain three foreground-subtracted WMAP CMB maps for Q, V, and 
W bands. The WMAP team also presented the Internal Linear Combination (ILC)
by computing a weighted combination of the maps that have been band-averaged 
within each of the five WMAP frequency bands, all smoothed to $1\degr$ 
resolution (Bennett et al. 2003b).
Tegmark et al. (2003) independently produced a foreground- and 
noise-cleaned WMAP map with W band resolution using a foreground-subtraction 
technique different from that of WMAP team. As compared to the combined Q, V,
and W band maps, Tegmark et al.'s cleaned map (TCM) has higher signal-to-noise 
ratio, and can be used as an independent probe of Gaussianity. 

These maps are stereographically projected on to a plane before the genus 
is measured. The stereographic mapping is conformal and locally preserves 
shapes of structures (see, e.g., Calabretta \& Greisen 2002, $\S5$). 
The final Q, V, and W band maps have resolution of $0\fdg60$, $0\fdg45$, 
and $0\fdg35$ FWHM, respectively, due to the additional smoothing during 
the projection. 
We mask 23.2\% of the sky that has significant foreground contamination from 
our Galaxy and point sources using the conservative Kp0 mask (Bennett et al. 
2003b). We also remove additional 0.3\% of the sky corresponding to the small
islands outside the north and south caps. The remaining survey area 
in the genus measurement is 76.5\% of the sky. 

\section{Genus Measurement}
We use the two-dimensional genus statistic introduced by Melott et al. (1989)
and Gott et al. (1990) as a quantitative measure of topology of the CMB 
anisotropy field. 
For the two-dimensional CMB anisotropy temperature field, the genus is the 
number of hot spots minus the number of cold spots. Equivalently, the genus
at a given threshold level $\nu$ is 
\begin{equation}
   g(\nu) = {1 \over {2\pi}} \int_C \kappa ds,
\end{equation}
where $\kappa$ is the signed curvature of the iso-temperature contours $C$, 
and the threshold level $\nu$ is the number of standard deviations from 
the mean. At a given threshold level, we measure the genus by integrating
the curvature along iso-temperature contours. The curvature is positive
(negative) if the interior of a contour has higher (lower) temperature than
the specified threshold level. Compared to the CONTOUR2D algorithm 
(Melott et al. 1989) that was used in Colley \& Gott (2003), this direct 
contour-integration method has an advantage that it can accurately calculate 
the genus when the survey region is enclosed by complicated boundaries (see
Fig. 4$a$ below). 

\begin{figure}
\includegraphics[width=84mm]{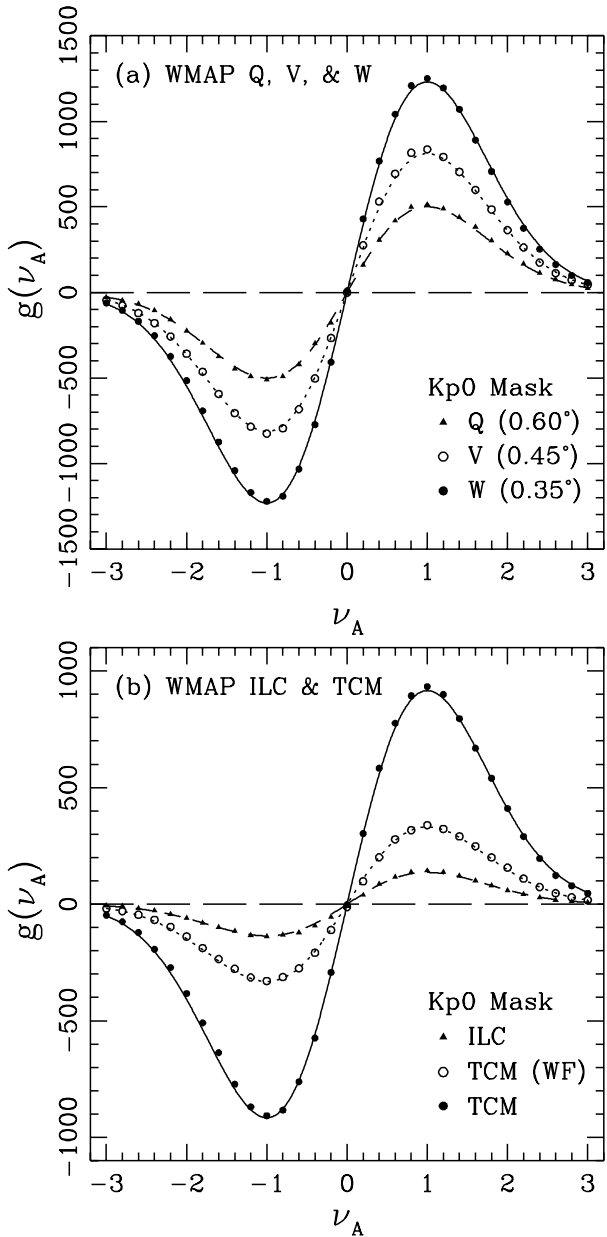}
\caption{Genus per steradian measured from ($a$) the foreground-subtracted
WMAP Q, V, and W band maps with angular resolution of $0\fdg60$, 
$0\fdg45$, and $0\fdg35$ FWHM, respectively, and from ($b$) ILC 
($\simeq 1\degr$), TCM, and Wiener-filtered TCM (WF, $0\fdg35$). 
Before the genus is measured, all maps are stereographically projected. 
The Kp0 mask is applied.
Curves in each panel show the functional forms expected for a random-phase
Gaussian field, $A \nu e^{-\nu^2 /2}$, and have been fitted to the measured
genus points by adjusting $A$.}
\end{figure}

\begin{figure}
\includegraphics[width=84mm]{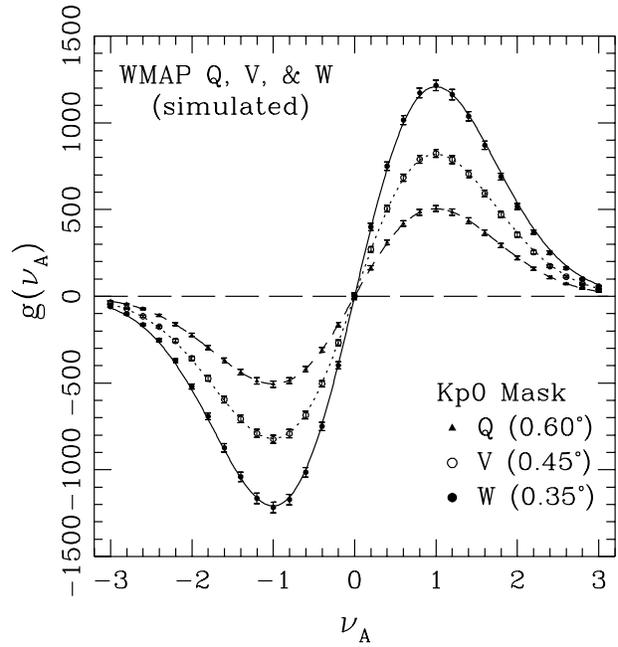}
\caption{Average genus per steradian measured from 500 WMAP Q (triangles), 
V (open circles), and W band (filled circles) mock observations. 
The error bars are given by the standard deviations of the genus values 
estimated from the mock observations, and the fitting curves are the functional
form expected for a random-phase Gaussian field 
(see the caption of Figure 1).}
\end{figure}

We present the genus curves as a function of the area fraction threshold level
$\nu_A$. The $\nu_A$ is defined to be the temperature threshold level at which 
the corresponding iso-temperature contours encloses a fraction of the survey
area equal to that at the temperature threshold level $\nu_A$ for a Gaussian 
field
\begin{equation}
   f_A = {1 \over {(2\pi)^{1/2}}} \int_{\nu_A}^{\infty} e^{-x^2 /2} dx .
\end{equation}
The $\nu_A=0$ level corresponds to the median temperature because 
this threshold level divides the map into high and low regions of equal area. 
Unlike the genus with the temperature threshold level, the genus with the area
fraction threshold level is less sensitive to the higher order information
(e.g., skewness) coming from the one-point distribution (Vogeley et al. 1994).
For each map we calculate area fraction threshold levels on a sphere 
with HEALPix format and with the same resolution as the stereographically 
projected map. This is because the stereographic mapping does not preserve 
the areas. 

For a two-dimensional random phase Gaussian field, the genus has a form of  
$g(\nu) = A \nu e^{-\nu^2 /2}$ (Gott et al. 1990). The amplitude $A$ is 
normalised so that $g(\nu)$ is the genus per steradian. 
Non-Gaussian feature in the CMB anisotropy will appear as deviations of 
the genus curve from this relation. Non-Gaussianity can shift the observed 
genus curve to the left or right directions, and also alter the amplitudes 
of the genus curve at positive and negative levels differently, 
causing $|g(\nu \approx -1)| \ne |g(\nu \approx +1)|$. 
We define shifts at negative and positive threshold levels, 
$\Delta\nu_{-/+}$, with respect to the Gaussian relation by minimizing
the $\chi^2$ between the genus points and the fitting function
\begin{equation}
   G = A_{-/+} \nu' e^{-\nu'^2 /2},
\end{equation}
where $\nu' = \nu - \Delta\nu_{-/+}$ and the fitting is performed over the 
range $-2.4 \le \nu \le -0.2$ and $0.2 \le \nu \le 2.4$, respectively.
Asymmetry parameter is defined as 
\begin{equation}
\Delta g \equiv 2 \left( { {A_{+} - A_{-}} \over {A_{+} + A_{-}}} 
   \right).
\end{equation}
Positive $\Delta g$ means that more hot spots are present than cold spots.
The Galactic foreground emission as well as point sources have significant 
effects on the genus curve even at high Galactic latitudes. The Galactic
foregrounds make the genus shifted to the left at all threshold levels 
while the radio point sources cause positive genus asymmetry 
(see Table 3 of PPR).
\begin{table*}
\centering
\begin{minipage}{110mm}
\caption{Genus-Related Statistics Measured from the WMAP CMB Maps}
\begin{tabular}{lrrrr}
\hline
  Maps &  $A$  & $\Delta\nu_{-}$  & $\Delta\nu_{+}$  &  $\Delta g$ \\
\hline
\multicolumn{5}{c}{Kp0 mask} \\
\hline
WMAP Q (Foreground-Subtracted)  &  $842$  & $+0.003$ & $+0.005$ & $+0.006$ \\
WMAP V (Foreground-Subtracted)  & $1361$  & $+0.008$ & $-0.015$ & $+0.017$ \\
WMAP W (Foreground-Subtracted)  & $2029$  & $+0.019$ & $-0.017$ & $+0.018$ \\
WMAP W (Foreground-Subtracted, North) & $2132$ & $+0.024$ & $-0.019$ & $-0.016$ \\ 
WMAP W (Foreground-Subtracted, South) & $1939$ & $+0.006$ & $-0.014$ & ${\bf+0.047}$ \\ 
\hline
ILC        & $ 231$ & $+0.042$ & $-0.002$ & $+0.017$ \\ 
TCM (WF)   & $ 546$ & $+0.019$ & $+0.012$ & $+0.029$ \\ 
TCM        & $1509$ & $+0.018$ & $-0.009$ & $+0.030$ \\ 
TCM (North)& $1611$ & $+0.015$ & $-0.006$ & $-0.006$ \\ 
TCM (South)& $1421$ & $+0.013$ & $-0.009$ & ${\bf+0.057}$ \\ 
\hline
\multicolumn{5}{c}{Kp0 mask \& $b=30\degr$ cut} \\
\hline
WMAP W        & $2031$ & $+0.015$ & $-0.022$ & $+0.025$ \\ 
WMAP W (North)& $2107$ & $+0.020$ & $-0.027$ & $+0.007$ \\ 
WMAP W (South)& $1960$ & $+0.011$ & $-0.015$ & ${\bf+0.045}$ \\ 
TCM        & $1523$ & $+0.013$ & $-0.016$ & $+0.020$ \\ 
TCM (North)& $1587$ & $+0.008$ & $-0.015$ & $+0.005$ \\ 
TCM (South)& $1464$ & $+0.019$ & $-0.013$ & ${\bf+0.047}$ \\ 
\hline
\multicolumn{5}{c}{Kp0 mask \& $b=40\degr$ cut} \\
\hline
WMAP W        & $1982$ & $+0.015$ & $-0.022$ & $+0.018$ \\ 
WMAP W (North)& $2038$ & $+0.014$ & $-0.023$ & $+0.005$ \\ 
WMAP W (South)& $1944$ & $+0.016$ & $-0.020$ & $+0.033$ \\ 
TCM        & $1486$ & $+0.017$ & $-0.027$ & $+0.016$ \\ 
TCM (North)& $1536$ & $+0.008$ & $-0.022$ & $+0.002$ \\ 
TCM (South)& $1456$ & $+0.026$ & $-0.024$ & $+0.033$ \\ 
\hline
\end{tabular}
\end{minipage}
\end{table*}

As shown in Figure 1$a$, the genus measured from the foreground-subtracted 
Q, V, and W band maps appear to have shapes consistent with Gaussian random
phase field. Compared to the Gaussian fitting curves, there is not any 
noticeable shift and asymmetry in the genus curves at all bands.
The genus measured from the ILC, TCM, and Wiener filtered TCM also show 
features similar to those of WMAP Q, V, and W band maps (Fig. 1$b$).
Table 1 summarises the genus-related statistics, namely the genus amplitude
($A$), the shift ($\Delta\nu_{-}$ and $\Delta\nu_{+}$), and the asymmetry 
($\Delta g$) parameters for the observed WMAP CMB maps of different bands.
The measured genus shifts at both negative and positive threshold levels 
as well as the genus asymmetry are very small.

\begin{table*}
\centering
\begin{minipage}{120mm}
\caption{Genus-Related Statistics Measured from the WMAP Gaussian Mock Maps}
\begin{tabular}{lcccc}
\hline
  Maps &  $A$  & $\Delta\nu_{-}$  & $\Delta\nu_{+}$  &  $\Delta g$ \\
\hline
\multicolumn{5}{c}{Kp0 mask} \\
\hline
WMAP Q  &$832\pm25$ & $+0.003\pm0.015$ & $-0.006\pm0.015$ & $-0.002\pm0.020$ \\ 
WMAP V &$1351\pm34$ & $+0.007\pm0.012$ & $-0.010\pm0.012$ & $-0.001\pm0.016$ \\ 
WMAP W &$1992\pm44$ & $+0.012\pm0.010$ & $-0.013\pm0.011$ & $-0.001\pm0.014$ \\ 
WMAP W (North) & $2002\pm64$ & $+0.011\pm0.014$ & $-0.013\pm0.014$ 
               & $-0.001\pm0.019$ \\ 
WMAP W (South) & $1986\pm59$ & $+0.012\pm0.014$ & $-0.013\pm0.015$ 
               & $-0.001\pm0.020$ \\ 
\hline
\multicolumn{5}{c}{Kp0 mask \& $b=30\degr$ cut} \\
\hline
WMAP W        & $1950\pm52$ & $+0.014\pm0.012$ & $-0.017\pm0.013$ 
              & $-0.001\pm0.017$ \\  
WMAP W (North)& $1957\pm75$ & $+0.014\pm0.018$ & $-0.016\pm0.018$ 
              & $-0.001\pm0.023$ \\ 
WMAP W (South)& $1947\pm72$ & $+0.014\pm0.018$ & $-0.016\pm0.019$ 
              & $-0.001\pm0.025$ \\ 
\hline
\end{tabular}
\end{minipage}
\end{table*}

To quantify the significance level for the non-Gaussianity from the measured
genus shift and asymmetry, we simulate 500 WMAP Gaussian CMB maps for each 
differencing assembly at each band. To make the genus amplitudes for the 
mock observations similar to those for the observed WMAP maps, 
we use the unbinned WMAP power spectrum (Spergel et al. 2003) at low $\ell$'s 
($\le 342$) and the CMBFAST-generated power spectrum (Seljak \& Zaldarriaga 
1996) that fits the WMAP power spectrum together with the CBI and ACBAR 
results at high $\ell$'s ($> 342$). 
For each mock observation, the same initial condition for $a_{\ell m}$'s 
is used for all frequency channels, where $a_{\ell m}$'s are coefficients 
in spherical harmonic expansion of the temperature anisotropy field.
During the map generation, the WMAP beam transfer function $B_\ell$ 
is used for each differencing assembly, and the instrument noise at each pixel 
is randomly drawn from the Gaussian distribution with variance of 
$\sigma_0^2/N_{\rm obs}$. The average genus measured from the 500 mock
WMAP observations are shown in Figure 2.

The genus-related statistics for the 500 mock observations are listed 
in Table 2, where the mean and standard deviation for each genus-related 
parameter are shown. 
Due to the complex noise property in the WMAP data, the genus shift parameters 
have means far deviating from zero, with $\Delta\nu_{-}$ and $\Delta\nu_{+}$ 
having opposite signs with each other.
Figure 3 shows the distribution of genus shift and asymmetry parameters 
drawn from the 500 WMAP W band Gaussian mock observations. 
The shift and asymmetry parameters well follow the Gaussian distribution.
Comparing the observed genus-related statistics with those for mock 
observations, we conclude that the WMAP temperature fluctuation is consistent
with the Gaussian field. This confirms the recent results of Komatsu et al.
(2003) and Colley \& Gott (2003).

\begin{figure}
\includegraphics[width=84mm]{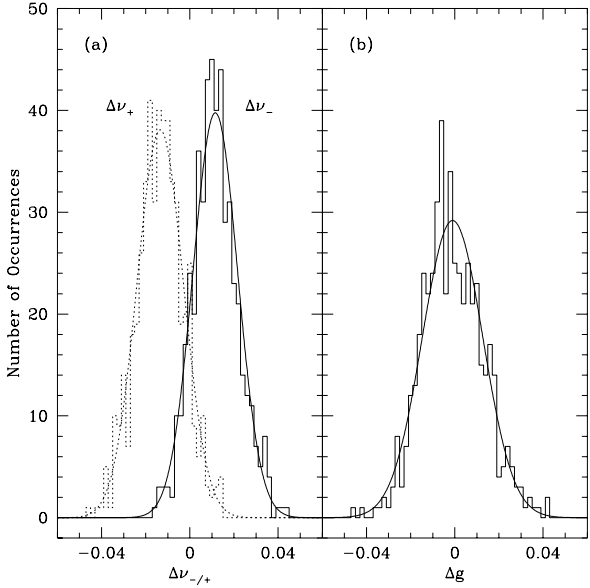}
\caption{Histograms of genus ($a$) shifts $\Delta\nu_{-}$ (solid), 
$\Delta\nu_{+}$ (dotted lines) and ($b$) asymmetry ($\Delta g$) parameters 
drawn from the 500 WMAP W band Gaussian mock observations. 
Gaussian functions with standard deviations given in Table 2 are used to
fit the histograms by adjusting the amplitudes. }
\end{figure}

The theoretical genus curve expected from the WMAP observation can be 
obtained by (Park et al. 1998)
\begin{equation}
   g(\nu) = { 1 \over 2(2\pi)^{3/2}} 
            {{\sum \ell(\ell+1)(2\ell+1) 
	    \left[ C_\ell W_\ell^2 + N_\ell \right] F_\ell^2 } \over 
	    {\sum (2\ell+1) 
	    \left[ C_\ell W_\ell^2 + N_\ell \right] F_\ell^2 }} 
	    \nu e^{-\nu^2 / 2},
\end{equation}
where $C_\ell$ and $N_\ell$ are WMAP CMB and noise power spectra, 
$W_\ell = B_\ell P_\ell$ is the window function that describes 
the combined smoothing effects of the beam ($B_\ell$) and the finite sky 
map pixel size ($P_\ell$), and $F_\ell$ is the additional smoothing 
filter which in our case has been used during the stereographic projection. 
We use $N_\ell$ as the average noise power spectrum derived from the noise 
power model given in Hinshaw et al. (2003). 
The theoretical genus amplitudes expected from the WMAP power spectrum 
are $A=866$, $1313$, and $2034$ for Q, V, and W bands, respectively, and
are very similar to those from mock observations.

\section{Genus in Northern and Southern Hemispheres}

As we have shown in the previous section, the CMB temperature fluctuation
measured by the WMAP appears to be consistent with Gaussian random phase 
field.
However, the temperature distributions of the Galactic northern and southern
hemispheres look very different with each other, 
especially due to the presence of large cold spots near the Galactic plane 
in the southern hemisphere (see Fig. 4). 
There are two big cold spots near $(\ell,b) \approx (330\degr, -10\degr)$ 
and $(200\degr,-20\degr)$. 
Figures 4$b$ and 4$c$ compare the temperature distributions of the ILC 
northern and southern hemispheres, showing that both are quite different 
from each other. 
\begin{figure*}
\begin{minipage}{140mm}
\includegraphics[width=140mm]{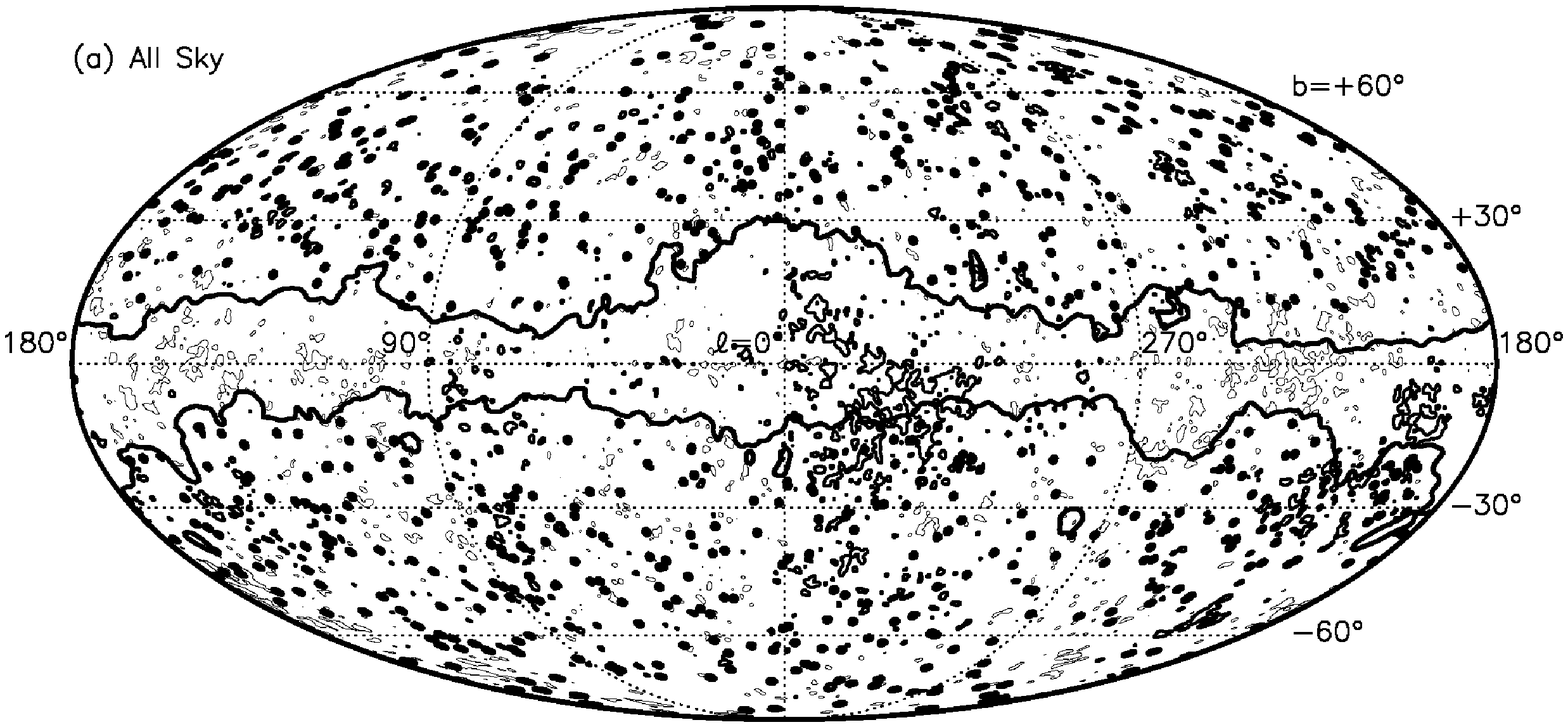} 
\includegraphics[width=140mm]{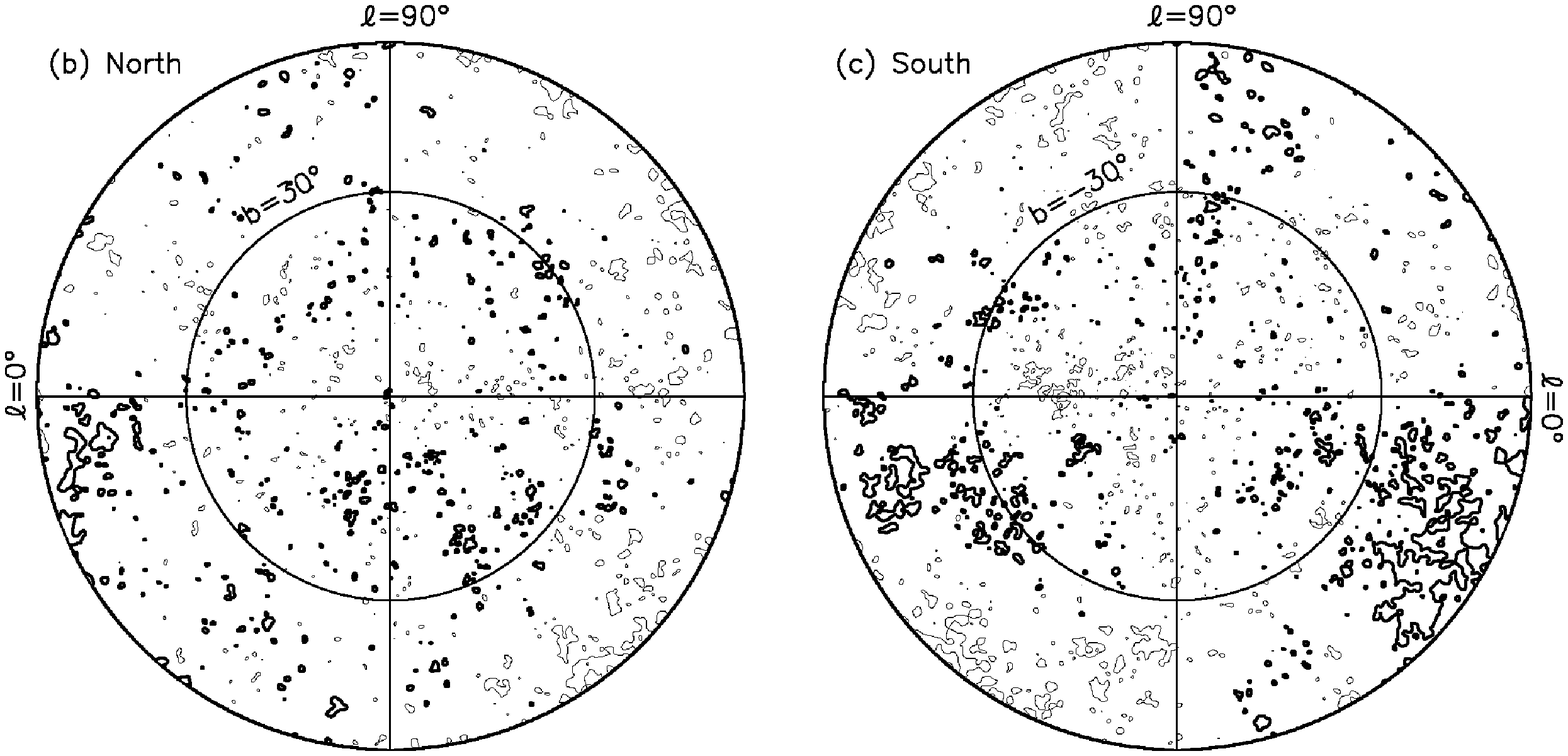} 
\caption{Iso-temperature contours of the ILC with about $1\degr$ resolution 
for ($a$) the whole sky, ($b$) the north, and ($c$) the south hemispheres 
at $\nu_A = -2.0$ (thick curves) and $\nu_A = +2.0$ (thin curves) threshold 
levels. 
Mollweide projection is used for the whole sky, where the Kp0 mask regions 
for removal of the strong Galactic emission and the bright radio point sources
are denoted as very thick solid curves and dots.
The maps for both hemispheres are stereographically projected. 
The temperature levels corresponding to $\nu_A = -2.0$ and $+2.0$ are 
$\Delta T = -144$ $\mu$K and $138$ $\mu$K, respectively, and have been 
estimated from the Kp0-masked ILC.}
\end{minipage} 
\end{figure*}
The northern hemisphere has very few big spots while the southern hemisphere 
has many. 

We calculate the genus separately from the northern and the southern 
parts of the WMAP maps and the TCM, and also measure the genus-related 
statistics. The results for the observed maps and for the 500 mock observations
are included in Tables 1 and 2, respectively. 
At the W band ($0\fdg35$ scale), the asymmetry parameter for the southern 
hemisphere is very large ($\Delta g = +0.047$), which is a non-Gaussian 
feature that is statistically significant at $2.4\sigma$ level. 
However, the $\Delta g$ for the north is consistent with zero. 
Such a large asymmetry does not appear in the lower resolution Q and V maps.

\begin{table} 
\centering
\caption{Genus Amplitude Differences between Northern and Southern Hemispheres
for WMAP Maps}
\begin{tabular}{llcr}
\hline 
Maps  &  Amp. Diff.  & Observed &  Simulated    \\
\hline
\multicolumn{4}{c}{Kp0 mask} \\
\hline
WMAP W &  $A_{-}^{\rm N} - A_{-}^{\rm S}$  & 255  &  $17\pm90$ \\
       &  $A_{+}^{\rm N} - A_{+}^{\rm S}$  & 131  &  $15\pm91$ \\
\hline
WMAP V &  $A_{-}^{\rm N} - A_{-}^{\rm S}$  & 140  &  $9\pm69$ \\
       &  $A_{+}^{\rm N} - A_{+}^{\rm S}$  & 142  &  $7\pm71$ \\
\hline
WMAP Q &  $A_{-}^{\rm N} - A_{-}^{\rm S}$  & 108  &  $6\pm51$ \\
       &  $A_{+}^{\rm N} - A_{+}^{\rm S}$  & 103  &  $5\pm53$ \\
\hline
TCM    &  $A_{-}^{\rm N} - A_{-}^{\rm S}$  & 235  &           \\
       &  $A_{+}^{\rm N} - A_{+}^{\rm S}$  & 145  &           \\
ILC    &  $A_{-}^{\rm N} - A_{-}^{\rm S}$  &  56  &           \\
       &  $A_{+}^{\rm N} - A_{+}^{\rm S}$  &  33  &           \\
\hline
\multicolumn{4}{c}{Kp0 mask \& $b=30\degr$ cut} \\
\hline
WMAP W &  $A_{-}^{\rm N} - A_{-}^{\rm S}$  & 184  &  $9\pm107$ \\
       &  $A_{+}^{\rm N} - A_{+}^{\rm S}$  & 110  &  $10\pm110$ \\
\hline
WMAP V &  $A_{-}^{\rm N} - A_{-}^{\rm S}$  & 107  &  $4\pm83$ \\
       &  $A_{+}^{\rm N} - A_{+}^{\rm S}$  & 125  &  $4\pm86$ \\
\hline
WMAP Q &  $A_{-}^{\rm N} - A_{-}^{\rm S}$  &  82  &  $3\pm62$ \\
       &  $A_{+}^{\rm N} - A_{+}^{\rm S}$  &  94  &  $2\pm64$ \\
\hline
TCM    &  $A_{-}^{\rm N} - A_{-}^{\rm S}$  & 152  &           \\
       &  $A_{+}^{\rm N} - A_{+}^{\rm S}$  &  93  &           \\
ILC    &  $A_{-}^{\rm N} - A_{-}^{\rm S}$  &  32  &           \\
       &  $A_{+}^{\rm N} - A_{+}^{\rm S}$  &  40  &           \\
\hline
\end{tabular}
\end{table}

Figure 5$a$ shows the genus for the northern and southern hemispheres measured 
from the W band data. Both genus curves have different genus amplitudes, 
especially at the negative threshold levels. 
We measure amplitude differences, $A_{-}^{\rm N} - A_{-}^{\rm S}$ and 
$A_{+}^{\rm N} - A_{+}^{\rm S}$, from the observed W band map (Kp0-masked)
and its 500 mock maps, and list the result in Table 3. 
Here $A_{-}^{\rm N(S)}$ is the amplitude obtained by fitting 
the genus points at $\nu_A < 0$ measured in the northern (southern) hemisphere
with a fitting function in equation (3), and likewise for $A_{+}^{\rm N(S)}$.
The $A_{-/+}^{\rm N} - A_{-/+}^{\rm S}$ have non-zero mean values 
even for the Gaussian mock observations.
The measured value $A_{-}^{\rm N} - A_{-}^{\rm S} = 255$ for the W band map 
is 2.6 times larger than the standard deviation of those expected 
from the simulated Gaussian observations ($[255-17]/90 \simeq 2.6$). 
On the other hand, $A_{+}^{\rm N} - A_{+}^{\rm S}$ has smaller value 
due to the large genus asymmetry ($\Delta g$) in the south.

We also have measured $A_{-/+}^{\rm N} - A_{-/+}^{\rm S}$ from the WMAP 
Q and V band maps, TCM, and ILC.
From Table 3, we find that the large genus amplitude difference between 
the north and the south also appear at both Q and V bands (Fig. 5$b$ and 5$c$),
deviating the Gaussian prediction with significance levels of about $2\sigma$ 
($A_{-/+}^{\rm N} - A_{-/+}^{\rm S} = 1.9\sigma$ -- $2.0\sigma$), 
which shows that this non-Gaussian feature does not have any frequency 
dependence.
We have computed a probability of finding such a significant deviation 
from the Gaussian prediction in the mock WMAP CMB fields with the Gaussian
statistics. The probability that the genus curves show such a large 
genus amplitude differences exceeding the observed values in Table 3
at all Q, V, and W bands is only 1.4\%.

\begin{figure*}
\begin{minipage}{160mm}
\includegraphics[width=160mm]{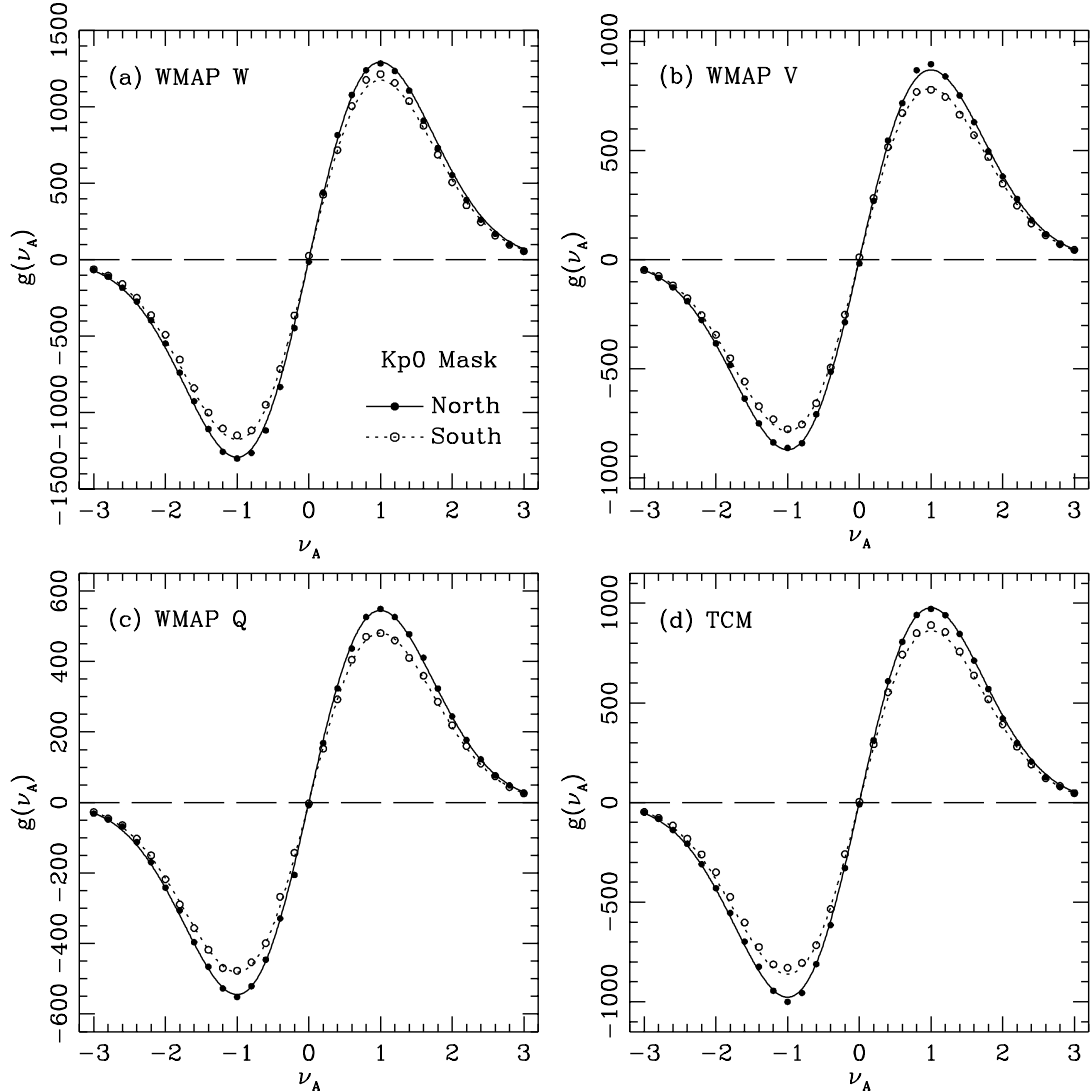} 
\caption{Genus per steradian measured from the WMAP maps and the TCM
for the northern (filled circles) and southern (open circles) hemispheres, 
together with Gaussian fitting curves (solid and dotted curves, respectively).}
\end{minipage} 
\end{figure*}

With a Galactic cut at $b=30\degr$, the large cold spots are removed 
(Fig. 4$c$). However, the trend of the genus amplitudes and asymmetries 
in the north and south are not significantly changed by the Galactic cuts 
(see Tables 1 and 3).
For the Galactic cut at $b=30\degr$, $A_{-}^{\rm N}-A_{-}^{\rm S}$ and 
$\Delta g$ still deviate from the Gaussian prediction at $1.6\sigma$ and 
$1.8\sigma$ levels, respectively. Since the genus amplitude significantly
decreases with higher Galactic cut, it is natural that the difference
between $A_{-/+}^{\rm N}$ and $A_{-/+}^{\rm S}$ should be smaller while 
its standard deviation in the mock observations becomes larger due to 
sample variance.
However, the probability that such genus amplitude differences are larger
than the measured values for $b=30\degr$ cut is still 4.6\%. 
All non-Gaussian features of the W band map described above are also seen 
in the TCM (Tables 1 and 3; Fig. 5$d$).

\section{Origin of the Non-Gaussianity}

Let us consider possible sources that may induce the non-Gaussian 
signatures shown in $\S4$. First, the foreground effect on the genus
may not be negligible, causing non-Gaussianity. 
The residual of the foregrounds after foreground-template correction 
is known to be very low. Bennett et al. (2003b) estimate that the residual 
Galactic contamination after the template correction is 2.2\%, 0.8\%, and 
$< 0.4$\% of the CMB power in Q, V, and W band, respectively, when the less 
conservative Kp2 mask is applied. Tegmark et al. (2003) also show 
that the foregrounds in their TCM are subdominant in all but the very 
innermost Galactic plane, for $\ell \la 100$. 
Furthermore, Galaxy foregrounds induce non-Gaussian feature by making 
the genus shifted to the left at all threshold levels.
As for the positive $\Delta g$ on the southern hemisphere in the high 
resolution W band data, it is more reliable that such non-Gaussianity 
is caused by point sources that have not been removed by the Kp0 mask 
because positive $\Delta g$ means that more hot spots are present than 
cold spots. Furthermore, another non-Gaussian signature, the large genus 
amplitude difference between the north and the south, does not show frequency 
dependence since all WMAP band maps (Q, V, and W), ILC and TCM show the similar 
non-Gaussian feature. It should be noted that the three independent methods 
of foreground-subtraction were applied to make the WMAP band maps, ILC, 
and TCM. 
However, we cannot conclude that the non-Gaussian signatures are free from 
the effect of foreground contamination. 
The $\Delta g$ and $A_{-/+}^N - A_{-/+}^S$ decrease as the higher Galactic 
cut is applied, though their dependence on the Galactic cut is weak.

Second, the non-Gaussianity can appear due to the Integrated Sachs-Wolfe (ISW)
and the Sunyaev-Zel'dovich (SZ) effects. 
The large-scale structures, which intersect the paths of CMB photons, 
can affect the primary CMB anisotropy. In a $\Lambda$-dominated CDM universe,
the CMB anisotropies are expected to be correlated with matter density
fluctuations at $z \la 2$ through the ISW effect (Crittenden \& Turok 1996).
Nolta et al. (2003) have detected cross-correlation of the NRAO VLA 
Sky Survey radio source catalog with the WMAP data. 
Boughn \& Crittenden (2003), Fosalba \& Gazta\~naga (2003), 
Fosalba, Gazta\~naga, \& Castander (2003), and Scranton et al. (2003) 
have reported that there exists a correlation between the WMAP temperature 
anisotropy and the galaxy distribution.
Diego, Silk, \& Sliwa (2003) estimate the cross-power spectrum 
between the WMAP CMB and the ROSAT X-ray maps, and did not find any significant 
correlation between two data sets. 
The SZ clusters appear in arcminute scales and the angular resolution 
and the sensitivity of WMAP are not ideal for detection of the typical 
SZ effect. However, if a large number of SZ clusters contribute to the CMB
map, their imprint can be statistically measurable.
Recently, the positive cross-correlation between the X-ray clusters and the 
WMAP map has been detected. 
Hern\'andez-Monteagudo \& Rubi\~no-Mart\'{\i}n (2003) claim their detection 
at $2\sigma$ -- $5\sigma$ level, which corresponds to the amplitude of 
typically 20 -- 30 $\mu$K in the WMAP map (see also Boughn \& Crittenden 2003; 
Fosalba \& Gazta\~naga 2003; Fosalba et al. 2003; Myers et al. 2003).

Third, the non-Gaussian signatures may be caused by the inhomogeneous photon
damping during the reionisation at the earlier epoch.
Reionisation damping depends on the total optical depth $\tau$ and the angular
scale subtended by the horizon at the last scattering surface defined during 
the reionisation epoch. The primary CMB anisotropy $\Delta T$ is damped to 
$\Delta T e^{-\tau}$ by the Thompson scattering with electrons, 
and the damping contribution to the anisotropy depends on scales.
The characteristic damping scale $\ell_r$ is given by 
$\ell_r = (1+z_r)^{1/2} (1+0.084\ln\Omega_0)-1$ (Hu \& White 1997; Griffiths, 
Barbosa, \& Liddle 1999), where $z_r$ denotes the reionisation epoch.  
For $z_r \approx 20$ and $\Omega_0 \approx 0.3$, the characteristic damping
scale is $\ell_r \approx 3$. 
Therefore the damping is also important at low $\ell$-modes.
For homogeneous reionisation, the reionisation process damps the CMB anisotropy
linearly and keeps the Gaussianity of the primordial fluctuations. 
However, if the reionisation is inhomogeneous, for example, 
due to the inhomogeneous ionisation fraction or the patchy reionisation, 
the CMB anisotropy of low $\ell$-modes can be partially contaminated 
by the inhomogeneous Thompson scattering, and becomes non-Gaussian. 
These low $\ell$-modes (quadrupole and octopole) happen to be aligned 
along the Galactic plane, make the biggest spots near the Galactic plane 
(Tegmark et al. 2003; de Oliveira-Costa et al. 2003), and may change  
the temperature distributions of the Galactic north and south hemispheres 
differently. 

Finally, the non-Gaussian signatures that have been detected may have 
the primordial origin. While the simple inflation predicts that the primordial
fluctuation field is Gaussian, the non-linear couplings between the inflaton
and the fluctuation fields or between the quantum and the classical fluctuation
fields can produce weakly non-Gaussian fluctuations (Komatsu \& Spergel 2001, 
references therein).
Multi-field inflationary models predict that the primordial metric fluctuations
do not need to obey Gaussian statistics (e.g., Bernardeau \& Uzan 2002). 
There are also many mechanisms that can generate non-Gaussian density 
perturbations during the inflation (e.g., Bartolo, Matarrese, \& Riotto 2002;
Dvali, Gruzinov, \& Zaldarriaga 2003). 

\section{Conclusions}

We have investigated the topology of CMB anisotropy from the WMAP 1-year maps 
by measuring the genus and its related statistics. 
The measured WMAP genus curves clearly deviate from the Gaussian prediction 
in two distinctive manners. 
First, the genus asymmetry parameter $\Delta g$ is positive on the Galactic
southern hemisphere in the WMAP W band map ($0\fdg35$ scale), 
a non-Gaussian feature that is significant at $2.4\sigma$ level, 
while $\Delta g$ on the Galactic northern hemisphere is consistent with zero. 
Second, the genus amplitude difference between the north and the south 
hemispheres ($A_{-/+}^{\rm N} - A_{-/+}^{\rm S}$) deviates from Gaussian 
prediction at $2.0\sigma$ -- $2.6\sigma$ levels at WMAP Q, V, and W bands. 
Compared to the 500 Gaussian mock observations, the observed genus amplitude
differences indicate that the Gaussianity of CMB anisotropy field is 
ruled out at about 99\% (95\%) confidence level when the Kp0 mask 
(Kp0 mask and $b=30\degr$ Galactic cut) is applied.
These non-Gaussian features have weak dependence on the Galactic cut.
Similar results based on the power spectrum analysis have also been 
reported by Eriksen et al. (2003).

To investigate the nature of the non-Gaussian features that have been detected,
we need CMB data with higher signal-to-noise ratio as well as the accurate 
models for the Galactic foregrounds. 
We wish that the next release of WMAP data or those of the future CMB 
experiments would enable us to resolve these problems.
As for the possibility of reionisation-induced non-Gaussianity at large scales,
we also need to study the topology of the CMB polarisation map 
(e.g., Park \& Park 2002).

\section*{Acknowledgments}
The author acknowledges valuable comments from Changbom Park and the anonymous
referee. This work was supported by the BK21 program of the Korean Government
and the Astrophysical Research Center for the Structure and Evolution of 
the Cosmos (ARCSEC) of the Korea Science and Engineering Foundation (KOSEF) 
through Science Research Center (SRC) program.
Some of the results in this paper have been derived using the HEALPix and 
CMBFAST packages.

\bsp

\label{lastpage}

\end{document}